\documentclass[12pt]{iopart}
\usepackage{amssymb}
\usepackage{graphicx}

\newcommand{\lmsp}{\gl_{\gs\gf}}
\newcommand{\eqref}[1]{(\ref{#1})}
\newcommand{\be}{\begin{eqnarray}}
\newcommand{\ee}{\end{eqnarray}}
\newcommand{\inv}{^{-1}}
\newcommand{\non}{\nonumber}
\renewcommand{\d}{\mathrm{d}}
\newcommand{\lh}{\left(}
\newcommand{\rh}{\right)}
\newcommand{\vc}[1]{{\mathbf{#1}}}
\newcommand{\siml}{\raisebox{-.1ex}{\renewcommand{\arraystretch}{0.3}
$\begin{array}{@{}c} \scriptstyle < \\ \scriptstyle \sim \end{array}$}}
\newcommand{\simg}{\raisebox{-.1ex}{\renewcommand{\arraystretch}{0.3}
$\begin{array}{@{}c} \scriptstyle > \\ \scriptstyle \sim \end{array}$}}
\DeclareMathSymbol{\mg}{\mathrel}{symbols}{"1D} 
\newcommand{\ml}{\ll}

\newcommand{\ga}{\alpha}
\newcommand{\gb}{\beta}
\renewcommand{\gg}{\gamma}
\newcommand{\gd}{\delta}
\renewcommand{\ge}{\epsilon}

\newcommand{\gk}{\kappa}
\newcommand{\gl}{\lambda}
\newcommand{\gm}{\mu}

\newcommand{\gr}{\rho}
\newcommand{\gs}{\sigma}

\newcommand{\gf}{\phi}
\newcommand{\gvf}{\varphi}

\newcommand{\gG}{\Gamma}

\newcommand{\gL}{\Lambda}

\newcommand{\cO}{{\mathcal O}}

\newcommand{\tn}{{\tilde n}}

\newcommand{\tge}{{\tilde\epsilon}}

\newcommand{\tget}{{\tilde\eta}}

\newcommand{\tgs}{{\tilde\sigma}}

\newcommand{\USP}{{\it Phys.\ Usp.\ }}

\newcommand{\JHEP}{{\it J.\ High\ Energy\ Phys.\ }}
\newcommand{\ApJS}{{\it Astrophys.\ J.\ Suppl.\ }}
\newcommand{\PRep}{{\it Phys.\ Rep.\ }}
\newcommand{\AnP}{{\it Annals Phys.\ }}
\newcommand{\NPPS}{{\it Nucl.\ Phys.\ Proc.\ Suppl.\ }}

\begin{document}
\title{Electroweak-scale inflation,
inflaton--Higgs mixing and the scalar spectral index}

\author{Bartjan van Tent\dag, Jan Smit\ddag~and Anders Tranberg\ddag}
\address{\dag\ DAMTP, Centre for Mathematical Sciences, University of
Cambridge,\\
Wilberforce Road, Cambridge CB3 0WA, United Kingdom.\\}
\address{\ddag\ Institute for Theoretical Physics, University of Amsterdam,\\
Valckenierstraat 65, 1018 XE Amsterdam, The Netherlands.}

\eads{\mailto{bvtent@damtp.cam.ac.uk}, \mailto{jsmit@science.uva.nl}, \
\mailto{anderst@science.uva.nl}}

\begin{abstract}

We construct a phenomenological model of electroweak-scale inflation that is in
accordance with recent cosmic microwave background observations by WMAP, while
setting the stage for a zero-temperature electroweak transition as assumed in
recent models of baryogenesis. We find that the scalar spectral index 
especially poses tight constraints for low-scale inflation models. The
inflaton--Higgs coupling leads to substantial mixing of the scalar degrees of
freedom. Two types of scalar particles emerge with decay widths similar to that
of the Standard Model Higgs particle.

\end{abstract}

\begin{indented}
\item[] \rm Keywords: inflation, physics of the early universe, 
baryon asymmetry, CMBR theory.
\end{indented}


\maketitle

\section{Introduction}

With the results from the WMAP mission \cite{WMAPparam,WMAPweb} it
has become relevant to critically review models of inflation,
especially with regard to the scalar spectral index. While clearly
still susceptible to improvement, the cosmic microwave background
(CMB) observations are accurate enough to rule out certain
(classes of) models, see e.g.\
\cite{WMAPinfl,Kinneyetal,LeachLiddle}. In this paper
we consider electroweak-scale inflation, which turns out to be indeed
tightly constrained by the spectral index.

The motivation for looking at electroweak-scale (i.e.\ of order
100~GeV) inflation is twofold. Firstly, it is interesting to see
if one can construct a working model of inflation with just
minimal extensions of the Standard Model (SM) of particle physics,
and to derive what kind of additional constraints such a coupling
to the SM puts on an inflation model.
The second (main) motivation has to do with baryogenesis, the
production of the observed baryon asymmetry in the universe. As
reviewed in
\cite{RubakovShaposhnikov}, all necessary ingredients for baryogenesis
(baryon number violation, C and CP violation, and non-equilibrium)
are present in the SM. This provides a strong motivation for
trying to construct a working model of electroweak baryogenesis.
However, the current lore is that in a standard finite-temperature
electroweak transition both the CP-violating and the
non-equilibrium effects are too small to be able to account for
the observed baryon asymmetry. These problems may be resolved in
the context of tachyonic preheating at the electroweak scale after
low-scale inflation \cite{KraussTrodden,GarciaBellidoetal,Felderetal,
Copelandetal,GarciaBellido,TranbergSmit}.
A tachyonic electroweak transition is strongly out of equilibrium,
and the fact that the process takes place at zero temperature
at the end of inflation may
maximize the effectiveness of CP violation
\cite{TalkSmit}.
In addition, the low reheating temperature prevents sphaleron wash-out of the
baryon number produced \cite{RubakovShaposhnikov}.

In this context it becomes
important to check that the models that combine low-scale
inflation with tachyonic electroweak preheating satisfy all the
new observational constraints from WMAP. Low-scale inflation has
been considered in many papers (see for instance
\cite{Knox,Kinney,Copelandetal,Germanetal,Lythrecent}).
In this paper we build in particular on
\cite{Copelandetal},
which was also motivated by the problem of electroweak
baryogenesis.
The main idea is that we have a kind of hybrid
inflation model \cite{ShafiVilenkin,Lindehybr},
in which inflation is driven by a nearly constant potential energy
of order (100~GeV)$^4$ \cite{KraussTrodden,GarciaBellidoetal},
while one field (the inflaton) slowly rolls down its
potential and the other field (the Higgs) is in a local minimum at zero.
Once the inflaton passes a critical value,
the local minimum for the Higgs field develops into a local maximum
and both fields roll down rapidly to
the absolute minimum at a non-zero value of the Higgs field,
thus breaking the electroweak symmetry. As was shown in
\cite{Copelandetal}, ordinary hybrid inflation models in which the
inflaton rolls from large field values towards zero are not viable
at low scales because of large quantum loop corrections. This
problem can be avoided in inverted hybrid inflation models, in
which the inflaton rolls away from zero and inflation takes place
at very small field values. Note that unlike standard hybrid
inflation, slow-roll inflation ends in this case before the
critical value is reached, instead of the end being caused by the
phase transition, so that the slow-roll inflation stage and the
phase transition can be considered as two separate processes.

The paper \cite{Copelandetal} was written before WMAP, and the
authors did not study the spectral index. As we will show in this
paper, their model gives a spectral index that is too low
according to WMAP. In \cite{Germanetal} somewhat more general
low-scale inflation models
were considered, although not from the point of view of
electroweak baryogenesis, but these models still
appear to be incompatible with WMAP.
We shall show that one can improve these models to obtain
a spectral index that lies comfortably within the WMAP confidence
levels.

At first sight it may seem that one can always fine-tune a model
with sufficient parameters to satisfy the constraints,
but this is not necessarily the case within a set of reasonable
rules. To formulate these rules we start from the point of view
that we are constructing a purely phenomenological model, a
minimal extension of the SM that introduces only one extra
inflaton field in order to describe the history of the universe
during and after inflation. We stress that there is nothing wrong
with fine-tuned parameters in a phenomenological model, as the
phenomenologically very successful SM shows. In incorporating a
slow-roll inflationary regime compatible with the CMB
measurements, one is led to an inflaton potential with
non-renormalizable couplings. To constrain this potential we
assume a polynomial approximation, such that there is only a
limited number of terms to be parametrized.

This leads to a tight fit when we also incorporate the scenario of
tachyonic electroweak baryogenesis, which requires that the
inflaton field ends up far from the slow-roll regime with a vacuum
expectation value similar to that of the Higgs field. The inflaton
is a gauge singlet and it couples only to the radial
(gauge-invariant) mode of the Higgs field. This coupling should
induce a sufficiently fast tachyonic electroweak transition to
make baryogenesis possible, without being unrealistically large.
It implies a considerable mixing between the inflaton and Higgs
modes, and the model predicts the existence of (only) {\em two}
scalar particle species with electroweak-scale masses. Up to
mixing-angle factors, their decay widths are similar to that of a
SM Higgs with the same mass. The model should therefore be
falsifiable by accelerator experiments, in particular with the LHC.

An important issue with any slow-roll inflation model is the
question whether the assumed flatness of the effective potential
is consistent with basic properties of quantum fields, with
`quantum corrections'. We investigate this by calculating one-loop
corrections to the effective potential. This exercise also led us
to a rough estimate of the scale at which the model may be
expected to break down because of its non-renormalizable and
strong couplings.

The outline of the paper is as follows. In section~\ref{Nk} we
first address the number of e-folds of inflation between horizon
crossing of a WMAP-observable scale and the end of inflation,
which number is crucial for the computation of CMB observables.
Contrary to the generic situation, there is little uncertainty
here because the (p)reheating of the universe and the onset of the
radiation-dominated era are reasonably well understood in this
model. Next, in section~\ref{model1}, we review the model of
\cite{Copelandetal}
and show that its spectral index disagrees with WMAP.
The implied inflaton--Higgs mixing is studied in
section~\ref{mixing}. We then show in the following section (plus
appendix~A) that, by adding two additional terms to the potential
and tuning the coupling parameters to a certain extent, values for
the scalar spectral index in agreement with WMAP can be obtained.
In section~\ref{loop} (plus appendix~B) we calculate one-loop
quantum corrections to the effective potential and study the
implications. Finally, section~\ref{concl} summarizes our
conclusions.

\section{Number of e-folds}
\label{Nk}

One of the most important differences between low-scale inflation
and `normal' inflation taking place around the GUT scale is that
the number of e-folds $N_k$ of inflation between horizon crossing
of the observationally relevant modes $k$ and the end of inflation
is much lower. An expression for $N_k$ is derived as follows
\cite{LiddleLyth} (see also the recent papers \cite{DodelsonHui,LiddleLeach}):
\begin{equation}
\fl
\frac{k}{a_0 H_0} = \frac{a_\mathrm{H} H_\mathrm{H}}{a_0 H_0}
= \frac{a_\mathrm{H}}{a_\mathrm{e}} \frac{a_\mathrm{e}}{a_\mathrm{reh}}
\frac{a_\mathrm{reh}}{a_\mathrm{eq}}
\frac{a_\mathrm{eq}}{a_0} \frac{H_\mathrm{H}}{H_0}
= e^{-N_k}
\lh \frac{\gr_\mathrm{eq}}{\gr_\mathrm{reh}} \rh^{1/4}
(1+z_\mathrm{eq})^{-1}\,
\frac{\gk \gr_\mathrm{H}^{1/2}}{\sqrt{3} H_0}.
\end{equation}
Here $a$ is the scale factor, $H\equiv\dot{a}/a$ the Hubble rate,
$z$ is the redshift, the subscripts $0$, H, e, reh and eq denote evaluation
now, at horizon crossing ($k=aH$), at the end of inflation, at the
end of reheating and at radiation--matter equality, respectively,
and $\gk$ is the inverse reduced Planck mass,
$\gk^2 \equiv 8\pi G$ ($\gk\inv = 2.436 \times 10^{18}$~GeV).
Here we used the fact that
$\gr \propto a^{-4}$ during radiation domination and the Friedmann
equation to rewrite $H_\mathrm{H}$. Furthermore we made use of the fact
that electroweak (p)reheating is nearly instantaneous on the
Hubble time scale at the end of inflation,
$a_\mathrm{e}/a_\mathrm{reh}= 1$, since its time
scale is of order of 1 GeV$^{-1}$ for the SM degrees of freedom
\cite{Skullerudetal},
whereas the Hubble time $H\inv$ at the end of electroweak-scale
inflation is of order
$10^{14}$ GeV$^{-1}$.
At radiation--matter equality the energy density $\gr_\mathrm{eq}$
is twice that in non-relativistic matter, $\gr_\mathrm{eq}
=2\Omega_\mathrm{m} (1+z_\mathrm{eq})^3\gr_{\mathrm{c}0}$, with 
$\gr_{\mathrm{c}0}$ the
critical density at present, $\gr_{\mathrm{c}0} = 3 \gk^{-2} H_0^2$.
Moreover, in our model $\gr_\mathrm{reh}$,
$\gr_\mathrm{e}$ and $\gr_\mathrm{H}$ are all practically equal, and so we find
\begin{equation}
N_k =
- \ln \lh \frac{k}{a_0 H_0} \rh +
\frac{1}{4} \ln \lh
\frac{2\Omega_\mathrm{m}}{3(1+z_\mathrm{eq})}\rh
+ \frac{1}{4} \ln\lh\frac{\gk^2\gr_\mathrm{H}}{H_0^2}\rh
= 23.8,
\end{equation}
where we used \cite{WMAPparam} $\Omega_\mathrm{m} = 0.29$,
$1+z_\mathrm{eq}= 3455$,
$H_0 = 73$~km/s~Mpc$\inv$,
the WMAP pivot scale of $k/a_0 = 0.05$~Mpc$\inv$ and an
inflationary energy scale of $\gr_\mathrm{H}^{1/4} = 100$~GeV. Hence,
this number of e-folds $N_k$ is much smaller than the 50--60 one
gets in the customary models where inflation takes place at much
higher energy scales.

As we will show below, the scalar spectral index $\tn \equiv n-1$ is
approximately inversely proportional to $N_k$. This means that the smaller
$N_k$ of low-scale inflation makes it more difficult to satisfy the WMAP
constraint that $\tn$ should be close to zero.
More precisely the constraints from WMAP (including CBI and ACBAR,
but no other experiments) for the amplitude
$|\gd_\vc{k}|^2$ and spectral index $\tn$ of the CMB power
spectrum are given by \cite{WMAPparam} (converted to our normalization, see the
definitions in \eqref{tn}):
\begin{equation}
|\gd_\vc{k}|^2 = (3.8 \pm 0.5) \times 10^{-10},
\qquad\qquad
\tn = -0.03 \pm 0.03.
\label{WMAPconstr}
\end{equation}

\section{Core model}
\label{model1}
\subsection{Model and WMAP constraints}

The model proposed in \cite{Copelandetal} contains a scalar field
$\gs$ (the inflaton) in addition to the SM Higgs field
$\gf$. The effective potential of the scalar fields has the
form\footnote{Here $\gf^2/2$ stands for $\gvf^{\dagger}\gvf$, with
$\gvf$ the usual complex SU(2) Higgs doublet of the SM.}
\begin{equation}
V(\sigma,\gf) = V_0 - \frac{1}{p} \ga_p \sigma^p + \frac{1}{q}
\ga_q \sigma^q - \frac{1}{2} \lmsp \sigma^2 \gf^2
+ \frac{1}{2} \gm^2 \gf^2 + \frac{1}{4} \gl_\gf
(\gf^2)^2,
\label{model}
\end{equation}
with integer $q>p>2$. The authors of \cite{Copelandetal} arrived
at the values
$p=5$, $q=6$. As we shall explain below, the value of $p$ is
fixed by matching the inflationary (small $\gs$) part of the
potential to the SM physics part where $\gs$ and $\gf$ are near
their vacuum expectation values. First we need to establish the
connection with the CMB data (\ref{WMAPconstr}).

We choose the inflationary energy scale $V_0^{1/4} = 100$ GeV.
This choice guarantees that after preheating and thermalization,
the temperature $T_\mathrm{reh}$ is substantially below the
electroweak crossover temperature $T_\mathrm{c}\simeq 70$ GeV
\cite{Tcross}, thereby avoiding sphaleron washout of the generated
baryon asymmetry.  The reheating temperature can be estimated as
$T_\mathrm{reh} = [30 V_0/(\pi^2 g_*)]^{1/4}\simeq 0.43\, V_0^{1/4}$,
with $g_* = 86.25$ the effective number of SM degrees of freedom
below the W~mass.\footnote{If there are in addition three
thermalized relativistic sterile neutrinos, $g_* =91.5$.}

Initial conditions are assumed such that the inflaton has a tiny
but non-zero value $\sigma_0 \siml 10^{-10}$~GeV. The Higgs field
is assumed to be in the ground state corresponding to this value
of $\gs$, i.e.\ $\gf=0$. Only when
$\sigma$ reaches the critical value $\sigma_\mathrm{c} =\gm/\sqrt{\lmsp}$, which
happens long after
inflation has ended in this model, will $\gf$ roll away from zero
and break the electroweak symmetry. This means that we can
consider the single-field slow-roll inflation stage and the phase
transition as two separate processes.

At the tiny values of $\gs$ relevant for inflation the $\sigma^q$
term in the potential is irrelevant. Taking only the first two
terms into account, the model can be solved analytically in the
slow-roll approximation in terms of the number of e-folds
$N$ since the beginning of inflation:
\be
\fl
\sigma_{,N} = - \frac{V_{,\sigma}}{3 H^2}
= \frac{\ga_p \sigma^{p-1}}{\gk^2 V_0}
\qquad \Rightarrow \qquad
\sigma(N) = \lh \sigma_0^{2-p} + (2-p) \frac{\ga_p}{\gk^2 V_0} \, N
\rh^{\frac{1}{2-p}},
\non \\
\fl
\tge = \frac{1}{2} \gk^2
\left(\sigma_{,N}\right)^2 = \frac{\ga_p^2}{2 \gk^2 V_0^2}
\, \sigma^{2p-2},
\qquad\qquad
\tget = \frac{\sigma_{,NN}}{\sigma_{,N}} - \tge = (p-1) \frac{\ga_p}{\gk^2 V_0}
\, \sigma^{p-2},
\label{oplmodel}
\ee
where $\tge \equiv -\dot{H}/H^2$ and
$\tget \equiv \ddot{\sigma}/(H\dot{\sigma})$
are the slow-roll functions, and we have used the fact that $\tge
\ml \tget$, because $\ga_p \sigma^p \ml V_0$ during inflation,
to neglect the $\tge$ term in the expression for $\tget$, as well
as to set $H^2=(\gk^2/3)V_0$. Defining $\tget=1$ as the end of
inflation, we can now compute the scalar amplitude and spectral
index to leading order in slow roll (see e.g.\
\cite{LiddleLyth,VanTent}):
\be
\fl
|\gd_\vc{k}|^2 = \frac{\gk^2}{50\pi^2} \frac{H_\mathrm{H}^2}{\tge_\mathrm{H}}
= \frac{1}{75\pi^2} \, \ga_p^{\frac{2}{p-2}} \lh \gk^2 V_0 \rh^{\frac{p-4}{p-2}}
\lh (p-1) + (p-2) N_k \rh^{\frac{2p-2}{p-2}},
\non \\
\fl
\tn = -2 \tget_\mathrm{H} - 4 \tge_\mathrm{H} = -2 \, 
\frac{(p-1)}{(p-1)+(p-2)N_k}
\approx - \frac{2}{N_k} \frac{p-1}{p-2},
\label{tn}
\ee
using again that $\tge_\mathrm{H} \ml \tget_\mathrm{H}$.
Hence we see that $\tn$ is indeed approximately inversely proportional to $N_k$.

From \eqref{tn} we see that the best (least negative) value one
can get is $\tn = -0.084$ (in the limit of large $p$). Actually
the situation is even worse than this, for two reasons. In the
first place we find in numerical studies that generically there
are about 2 more e-folds of inflation after $\tget=1$ has been
reached, which means that one should use $N_k \simeq 22$ instead
of
$23.8$. Secondly, one cannot take an arbitrarily large $p$, because, as we will
show below, $p>5$ is not compatible with the constraints from the
Higgs sector. This means that the upper limit is actually
$\tn = -0.114$ in this model, which is $3 \sigma$ away from the WMAP
result \eqref{WMAPconstr}.

\begin{table}
\begin{center}
\begin{tabular}{l|llll}
& \multicolumn{1}{c}{$\ga_p$} & \multicolumn{1}{c}{$\sigma_\mathrm{e}$ (GeV)}
& \multicolumn{1}{c}{$\sigma_\mathrm{H}$ (GeV)} 
& \multicolumn{1}{c}{$\tge_\mathrm{e}$}\\
\hline
$p=4$ & $2.71 \times 10^{-12}$ & $1.4 \times 10^{-9}$ & $3.6 \times 10^{-10}$
& $1.9 \times 10^{-56}$\\
$p=5$ & $1.51 \times 10^{-3}$~GeV$\inv$ & $1.4 \times 10^{-9}$
& $5.4 \times 10^{-10}$ & $1.0 \times 10^{-56}$\\
$p=6$ & $6.75 \times 10^{+5}$~GeV$^{-2}$ & $1.5 \times 10^{-9}$
& $7.2 \times 10^{-10}$ & $7.5 \times 10^{-57}$
\end{tabular}
\end{center}
\caption[infltab]{Some quantities in the inflation model with potential
$V=V_0-\frac{1}{p}\ga_p \sigma^p$ with $V_0=(100$~GeV$)^4$ for $p=4,5,6$,
as follow from applying the WMAP amplitude constraint for the mode that
crossed the horizon at $N_\mathrm{H}$, defined as $N_k=22$ e-folds before 
the end of inflation $N_\mathrm{e}$.}
\label{infltab}
\end{table}

The coefficient $\ga_p$ is determined by fitting the amplitude
\eqref{tn} to the WMAP value \eqref{WMAPconstr}. Results are given
in table~\ref{infltab} for the cases of $p=4,5,6$. From this we
can derive at which value $\sigma_\mathrm{e}$ of the inflaton
slow-roll inflation ends ($\tget=1$), and at which value
$\sigma_\mathrm{H}$ horizon crossing of the scale under consideration
occurs, see the table. Note that $\tge$ at the end of inflation,
also given in the table, is still tiny.

\subsection{Higgs sector}
\label{Higgssector}

Next we look at the matching to the Higgs sector, i.e.\ we
consider the stage after inflation when $\sigma$ has grown larger
than $\sigma_\mathrm{c}$ and
$\gf$ is no longer zero. Here we have the following constraints.
Denoting the value of the fields in the absolute minimum by
$v_\sigma$ and $v_\gf$, respectively, we have
firstly the two conditions that the first derivatives of the
potential with respect to
$\sigma$ and $\gf$ in the point $(v_\sigma,v_\gf)$ vanish.
Secondly, we demand that the second derivative with respect to
$\gf$ in  this point is equal to the the diagonal Higgs mass
squared, $m_\gf^2$. The actual particle masses are to be obtained
from a diagonalization of the
$\gs$--$\gf$ mass matrix, which we will discuss in
section~\ref{mixing}. Thirdly, there is the condition that
$V(v_\sigma,v_\gf)=0$, so there is no residual vacuum energy
(or it is at least negligible compared to the electroweak
scale).

Taking $v_\gf$, $m_\gf$ and
$\gm$ as additional input parameters, these four constraints lead
to expressions for $\lmsp$, $\gl_\gf$, $\ga_q$ and $v_\sigma$:
\begin{equation}
\fl
\lmsp = \lh \gm^2 + \frac{1}{2} m_\gf^2 \rh v_\sigma^{-2},
\qquad
\gl_\gf = \frac{m_\gf^2}{2 v_\gf^2},
\qquad
\ga_q = \lh \gm^2 + \frac{1}{2} m_\gf^2 \rh v_\gf^2 v_\sigma^{-q}
+ \ga_p v_\sigma^{p-q}.
\label{method1}
\end{equation}
The vacuum expectation value of the inflaton, $v_\sigma$, cannot be given
analytically, but can easily be computed numerically from the condition that
$V(v_\sigma,v_\gf)=0$, given the above relations. We take
\begin{equation}
v_\gf = 246\;\mbox{GeV},
\qquad\qquad
m_\gf = 200\;\mbox{GeV},
\qquad\qquad
\gm = 100\;\mbox{GeV},
\label{inputs}
\end{equation}
where $v_\gf$ is the usual SM value. The value for the Higgs mass
is of course not known yet, but we require that the eigenvalues of
the $\gs$--$\gf$ mass matrix are above the current lower bound on
the Higgs particle mass of 114~GeV
\cite{pdg}. It turns out that to satisfy this lower bound we need
to choose a much larger value for the diagonal Higgs mass; 200 GeV
will be sufficient.

A further condition comes from baryogenesis: the inflaton--Higgs
coupling should be large enough that the tachyonic transition is
sufficiently out of equilibrium. The rate of change of the
effective Higgs mass squared when $\gs$ crosses 
$\gs_\mathrm{c} = \mu/\sqrt{\lmsp}$ is
determined by a dimensionless velocity parameter $u$:
\begin{equation}
u\equiv \frac{1}{2\mu^3}\, \frac{\partial (\mu^2 - \lmsp \gs^2)}{\partial t}
= - \frac{\sqrt{\lmsp}}{\mu^2}\, \frac{\partial \gs}{\partial t},
\qquad
t=t_\mathrm{c},
\end{equation}
where $t_\mathrm{c}$ is the time when $\gs = \gs_\mathrm{c}$. We have used 
$\mu$ to set the scale. The value of this mass is not critical here
(although it may influence the precise value of the baryon asymmetry generated) 
and we have chosen $\mu=100$~GeV. We shall assume that
$u \simg 0.15$ for sufficient baryogenesis.\footnote{Our
$u$ is equivalent to the velocity $V$ of \cite{GBetal2,GarciaBellido}.}
The value of $u$ can be estimated from energy conservation
\cite{Copelandetal}:
\be
\fl
\frac{1}{2} \left(\frac{\partial \gs}{\partial t}\right)^2 =
V_0 - V(\gs_\mathrm{c},0) \equiv c V_0,
\qquad
0<c<1
\qquad \Rightarrow \qquad \lmsp = \frac{\mu^4 }{2 c V_0}\, u^2.
\label{uestimate}
\ee
Requiring $u \simg 0.15$ means that $\lmsp \simg 0.01/c > 0.01$
(for our choices $\mu =100$~GeV and $V_0^{1/4} = 100$~GeV).
On the other hand,
$\lmsp$ should not be too large in order to have radiative corrections
under control, say $\lmsp < 1$.

From the expressions given in
\eqref{method1} and \eqref{inputs} we find the results given in
table~\ref{Higgstab}. Now we can draw the conclusion alluded to
before: $p=5$ is the only value that satisfies all constraints.
For $p\leq 4$ the coupling $\lmsp$ between the inflaton and the
Higgs is too small for baryogenesis, while this case is also worse
for the scalar spectral index $\tn$. For $p\geq 6$, $\lmsp$ is much
too large, making quantum corrections uncontrollable. Note that
even the extreme $p=6$ case with $q=7$,
$\gm=1$~GeV and $m_\gf=132$~GeV (which means that the smallest eigenvalue 
of the $\gs$--$\gf$ mass matrix is 115~GeV) gives a $\lmsp$ that is still much 
too large ($\lmsp=625$), so that this conclusion does not depend on our 
parameter choices. Hence, we really cannot go beyond $p=5$ when we try to 
maximize the spectral index.

\begin{table}
\begin{center}
\begin{tabular}{c|cccc}
& $\lmsp$ & $\gl_\gf$ & $\ga_q$ & $v_\sigma$ (GeV)\\
\hline
$p=4$, $q=6$ & $1.4 \times 10^{-6}$ & 0.33 & $3.2\times 10^{-22}$~GeV$^{-2}$
& $1.45\times 10^{5}$ \\
$p=5$, $q=6$ & 0.36 & 0.33 & $8.4 \times 10^{-6}$~GeV$^{-2}$ & 288\\
$p=6$, $q=7$ & $2.0 \times 10^3$ & 0.33 & $3.0 \times 10^{+5}$~GeV$^{-3}$
& 3.90\\
$p=6$, $q=8$ & $3.1 \times 10^3$ & 0.33 & $2.9 \times 10^{+5}$~GeV$^{-4}$ & 3.09
\end{tabular}
\end{center}
\caption[Higgstab]{Values for the parameters and inflaton vacuum expectation
value $v_\sigma$ in the full model \eqref{model} for different
choices of
$(p,q)$, following from constraints on the Higgs sector of the model, as
explained in the main text, using the values given in table~\ref{infltab} and
equation~\eqref{inputs}.}
\label{Higgstab}
\end{table}

The basic reason for the limit on $p$ is the fact that
$p>5$ implies a rapid turning down of the potential as $\gs$
increases and, together with the low value $V_0^{1/4} = 100$~GeV,
this leads to a very small $v_\gs$; this in turn requires a very
large $\lmsp$ in order to obtain the required value
$\lmsp v_\gs^2 - \mu^2 = m_\gf^2/2$.

We end this section by noting that for the $p=5$, $q=6$ case,
$\tilde n = -0.114$; furthermore $\sigma_\mathrm{c} = 166$~GeV,
which satisfies $\sigma_\mathrm{e} \ml \sigma_\mathrm{c} < v_\sigma$ as we
assumed, and $c=0.09$ (cf.\ (\ref{uestimate})).
Note that $\lmsp=0.36$ satisfies the requirement $\lmsp > 0.01/c$.
The energies corresponding to the values of
$\ga_5$ and $\ga_6$ are somewhat above the electroweak scale:
$1/\ga_5 = 661$ GeV, $1/\sqrt{\ga_6} = 344$ GeV.

\section{Inflaton--Higgs mixing}
\label{mixing}

In this section we take a first look at the possibility of testing
the model with accelerator experiments. The shape of the potential
near its absolute minimum determines the particle masses and
interactions that we can measure in the laboratory. The inflaton
and Higgs fields have the same quantum numbers after electroweak
symmetry breaking, and consequently the emerging particles
correspond to a mixture of the two fields. The particle masses are
given by the eigenvalues of the
$\gs$--$\gf$ mass matrix,
\begin{equation}
\left(\begin{array}{cc}V_{,\gs\gs}&V_{,\gs\gf}\\V_{,\gf\gs}&V_{,\gf\gf}
\end{array}\right),\qquad \gs=v_\gs,\;\; \gf=v_\gf,
\end{equation}
and for the core model with $p=5$, $q=6$ they are given by
$m_1 = 385$ GeV, $m_2 = 125$ GeV,
safely above the current experimental lower bound for
the Higgs mass of 114~GeV. The mixing angle defined by
\be
\gf_1 = \gs \cos \xi - \gf \sin\xi,
\qquad
\gf_2 = \gs \sin\xi + \gf\cos\xi,
\ee
where $\gf_1$ and $\gf_2$ are the mass eigenmodes, is given by
$\sin\xi =0.43$, $\cos\xi = 0.90$, or $\xi = 0.44$.

Because $m_1 > 2 m_2$, the heavier particle can decay into two
lighter ones and it is of interest to calculate the decay rate
(see e.g.\ \cite{DeWit}):
\be
\gG_{1\rightarrow 2+2} = \frac{\gg_{122}^2\sqrt{m_1^2 - 4 m_2^2}}{32\pi m_1^2},
\ee
with $\gg_{122}$ the three-point coupling:
\be
\gg_{122} = \frac{\partial^3 V}{\partial\gf_1\partial\gf_2\partial\gf_2},
\qquad
\gs=v_\gs,\;\;\;\gf=v_\gf.
\label{deflm122}
\ee
We find $\gg_{122} = 53$~GeV and a decay rate
$\gG_{1\rightarrow 2+2} = 56$~MeV. However, the mixing into the Higgs
proportional to $\sin\xi$ facilitates decays into the modes
allowed for a SM Higgs particle with mass $m_1$, e.g.\ the decay
into two W bosons, with a much larger rate reaching 100~GeV for a
mass
$\approx 500$~GeV \cite{pdg}. Since the latter decay mode is
forbidden for the lower mass particle 2, it decays predominantly
via the $b\bar b$ channel with a much smaller rate of order 10~MeV
\cite{pdg}. Hence the branching ratio via $1\rightarrow 2+2$ will be very small.
So the model predicts
\be
\Gamma_{1\rightarrow\mathrm{X}} \simeq \sin^2\xi\; 
\Gamma_{\mathrm{H}\rightarrow\mathrm{X}},
\qquad
m_\gf = m_1,
\\
\Gamma_{2\rightarrow\mathrm{X}} \simeq \cos^2\xi\; 
\Gamma_{\mathrm{H}\rightarrow\mathrm{X}},
\qquad
m_\gf = m_2,
\ee
for a dominant decay mode X of the SM Higgs particle.

Let us conclude this section by discussing the number of
parameters of the model. Consider coupling the gauge-singlet
inflaton field to the SM Higgs field in a renormalizable fashion.
To start out with this would introduce three new parameters beyond
those of the SM:
\be
m_\gs^2 \equiv V_{,\, \gs\gs}(v_\gs,v_\gf),
\qquad\qquad
\gl_\gs \equiv \frac{1}{6} \, V_{,\,\gs\gs\gs\gs}(v_\gs,v_\gf)
\label{deflmsg}
\ee
and $\lmsp$. Lacking the symmetry $\gs\rightarrow -\gs$, we should
also include $V_{,\, \gs\gs\gs}(v_\gs,v_\gf)$ and
$V_{,\, \gs\gf\gf}(v_\gs,v_\gf)$.
The linear term in the expansion around the minimum vanishes since
$V_{,\,\gs}(v_\gs,v_\gf)=0$, which determines $v_\gs$. We should
also keep in mind the cosmological constant, which is needed to
control the energy density of the vacuum (approximated by zero in
this paper) and which is usually not included in the parameter set
of the SM. So within the renormalizable class of models we should
not be surprised to find seven new parameters. The scenario for
baryogenesis suggests the introduction of two more parameters in
order to be able to adjust the height and the first derivative of
the potential at the spinodal point
$\gs_\mathrm{c}$: $V(\gs_\mathrm{c},0)$ and 
$V_{,\,\gs}(\gs_\mathrm{c},0)$ (where $\gs_\mathrm{c}$
is determined by $V_{,\, \gf\gf}(\gs_\mathrm{c},0)=0$). These two
conditions suggest that we need more freedom in the potential,
e.g.\ provided by the coefficients of non-renormalizable $\gs^5$
and $\gs^6$ terms.

In the core model all these parameters are basically set by the
inflationary parameters $V_0$ and $\ga_p$, except for the
cosmological constant which is controlled by $\ga_q$. It would of
course be surprising if the parameters thus obtained were just
right for the phenomenology in the electroweak domain. To
influence their values (e.g.\ to control $m_\gs^2$ and hence also
$m_1^2$) we would have to add further terms to the potential in
the electroweak regime. For the case $p=5$, $q=6$ these would be
$\gs^7$, $\gs^8$, \ldots. Experiment will inform
us of the need of such terms. Moreover, in the inflationary domain
there are terms that have been arbitrarily set to zero in the
potential of the core model with $p=5$, namely the $\gs^2$,
$\gs^3$ and $\gs^4$ terms. These will be considered in the next
section.

\section{Improving the core model}
\subsection{Adding a quadratic term}

To bring the scalar spectral index closer to the central WMAP
value we add a negative quadratic mass term for the inflaton to
the potential, as in the class of models studied in
\cite{Germanetal}:
\begin{equation}
V(\sigma,\gf) = V_{\mathrm{core}}(\sigma,\gf) - \frac{1}{2} \ga_2 \sigma^2,
\label{model2}
\end{equation}
with $V_{\mathrm{core}}(\sigma,\gf)$ defined in \eqref{model}.
In this case the slow-roll
system (considering only
$V_0-\frac{1}{2}\ga_2\sigma^2-\frac{1}{p}\ga_p\sigma^p$)
can still be solved analytically:
\be
\fl \sigma(N) = \left[ -\frac{\ga_p}{\ga_2} + \lh \sigma_0^{2-p}
+ \frac{\ga_p}{\ga_2} \rh
\exp \lh (2-p)\frac{\ga_2}{\gk^2 V_0} \, N \rh \right]^{\frac{1}{2-p}},
\non\\
\fl \tge = \frac{\ga_2^2}{2 \gk^2 V_0^2} \, \sigma^2 \lh 1 
+ \frac{\ga_p}{\ga_2} \, \sigma^{p-2} \rh^2,
\qquad\qquad
\tget = \frac{\ga_2}{\gk^2 V_0} \lh 1 + (p-1) \frac{\ga_p}{\ga_2}
\, \sigma^{p-2} \rh,
\ee
which agrees with \eqref{oplmodel} in the limit that $\ga_2 \rightarrow 0$.

\begin{figure}
\begin{center}
\includegraphics[width=10cm]{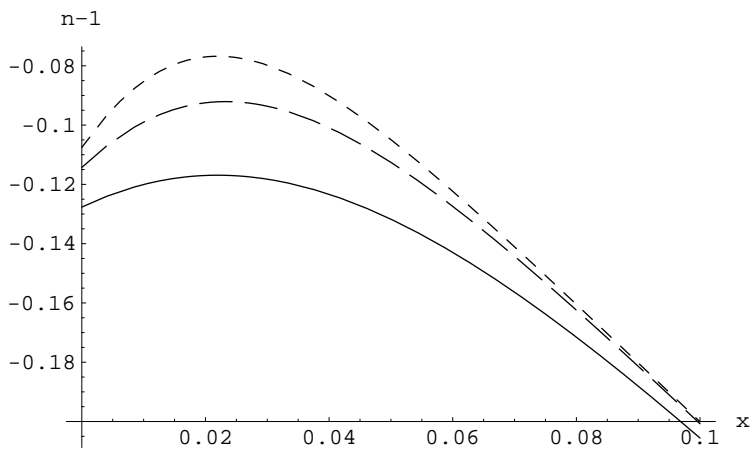}
\end{center}
\caption{\label{nvsx}The scalar spectral index $\tn \equiv n-1$ as a function
of the dimensionless mass parameter $x \equiv \ga_2/(\gk^2 V_0)$ in
the model
\eqref{model2}, as given in equation \eqref{tnx}, with $N_k=22$ and plotted for
$p=4$ (solid), $p=5$ (long dashes) and $p=6$ (short dashes).}
\end{figure}

In a derivation analogous to the one in the previous section, we find for the
spectral index the following expression:
\begin{equation}
\fl \tn = -2 x \, \frac{(p-2)(1-x) + (1+(p-2)x) e^{(p-2) x \, N_k}}
{(x-1)+(1+(p-2)x) e^{(p-2) x \, N_k}},
\qquad\mbox{with }
x \equiv \frac{\ga_2}{\gk^2 V_0}.
\label{tnx}
\end{equation}
The curve $\tn(x)$ is plotted in figure~\ref{nvsx} for $p=4,5,6$
and $N_k=22$. We see that the  situation can be somewhat improved
compared to the massless ($x=0$) case by  choosing $\ga_2$ such that
$x \approx 0.023$, i.e.\ $\sqrt{\ga_2} = 6.3 \times 10^{-16}$~GeV, which for
$p=5$ changes $\tn$ from $-0.114$ to $-0.092$. However, this is
still a $2 \sigma$ deviation from the central WMAP result. Note
that the added quadratic term is completely negligible for $\sigma
\sim v_\sigma$, and its only effect for the Higgs sector comes
from changing the value of $\ga_p$ that follows from the WMAP
amplitude (to $\ga_5 = 5.0 \times 10^{-3}$~GeV$\inv$ for the case
$p=5$). This does not change the conclusion about $p>5$ being
excluded.

\subsection{\ldots and a quartic term}
\label{quartic}

Including in addition a quartic term for the inflaton
allows for more variation in the spectral index:
\begin{equation}
V(\sigma,\gf) = V_{\mathrm{core}}(\sigma,\gf) - \frac{1}{2} \ga_2 \sigma^2
+ \frac{1}{4} \ga_4 \sigma^4,
\end{equation}
with $V_\mathrm{core}(\sigma,\gf)$ defined in \eqref{model}, where we have now
fixed $p=5$, $q=6$. This means that for the inflationary part we consider the
potential
\begin{equation}
V_\mathrm{infl}(\sigma) = V_0 - \frac{1}{2} \ga_2 \sigma^2
+ \frac{1}{4} \ga_4 \sigma^4 - \frac{1}{5} \ga_5 \sigma^5.
\label{Vinfl}
\end{equation}
For the analysis it turns out to be useful to make the following definitions:
\be
R \equiv \frac{27 \ga_2 \ga_5^2}{\ga_4^3},
\qquad\qquad
x \equiv \frac{\ga_2}{\gk^2 V_0},
\qquad\qquad
\tgs \equiv \frac{3 \ga_5}{\ga_4} \, \gs
\label{defsRx}
\ee
(this is the same definition for $x$ as in the previous subsection).
Assuming that all $\ga$'s are positive
there is always a maximum at
$\gs=0$, and for $\gs > 0$ there are three different cases
depending on the value of $R$. For $R > 27/4$ the potential has a
negative second derivative for all positive $\gs$, for $4 < R <
27/4$ the second derivative changes sign from negative to positive
and back to negative, and for $R < 4$ there is an additional
minimum--maximum pair. For $R = 4$ the potential has a flat
plateau around $\tgs = 2$. We restrict ourselves to the region
with $R > 4$, because an additional minimum would stop the
inflaton from ever reaching the critical value
$\sigma_\mathrm{c}$ (ignoring quantum tunnelling which we do not consider 
in this paper) and the electroweak symmetry would not be broken (nor would
inflation end at all). In the full model there is of course no
drop off to minus infinity, because after inflation the
$\frac{1}{6}\ga_6\sigma^6$ term will come into play to create the
absolute minimum at $v_\sigma$.

Using the slow-roll equation of motion, we can derive expressions for
$\tge(\gs)$ and $\tget(\gs)$ in this model as well, leading to the following
results for the amplitude and spectral index, still in terms of the inflaton
field:
\be
\fl
|\gd_\vc{k}|^2 = \frac{\ga_4}{225\pi^2} \lh \frac{R}{x} \rh^3
\lh R \tgs_\mathrm{H} - 3 \tgs_\mathrm{H}^3 + \tgs_\mathrm{H}^4 \rh^{-2},
\qquad
\tn = - \frac{2x}{R} \lh R - 9 \tgs_\mathrm{H}^2 + 4 \tgs_\mathrm{H}^3 \rh.
\label{amplitudeandslope}
\ee
To determine $\tgs_\mathrm{H}$ we need to solve the equation of motion.
Although it can be integrated analytically to give $N(\tgs)$, the
inversion to obtain $\tgs(N)$ has to be carried out numerically.
The horizon-crossing value $\tgs_\mathrm{H}$ is then the field value $22$
e-folds before the end of inflation, which is defined by the
relation $\tget=1$. The analytical results used in this procedure
are given in appendix~A.

\begin{figure}
\begin{center}
\includegraphics[width=10cm]{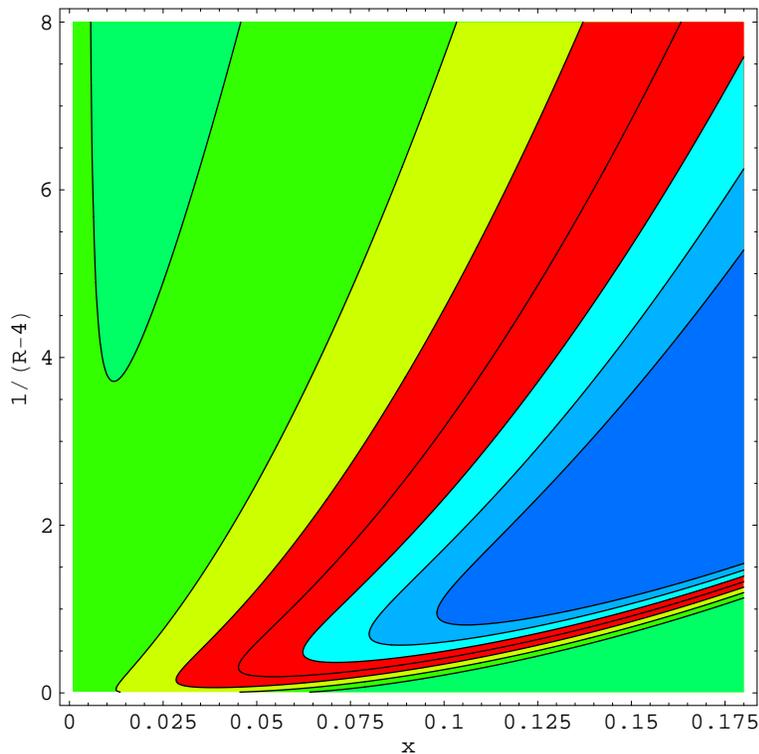}
\end{center}
\caption{\label{contn}Contourplot for $\tn \equiv n-1$
as a function of the parameters $x$ and $(R-4)^{-1}$ (defined in
\eqref{defsRx}) in the model \eqref{Vinfl} for the mode that left the horizon
$N_k=22$ e-folds before the end of inflation.
The contours are (from outside to inside) for $\tn = -0.12, -0.09,
-0.06, -0.03, 0.00, +0.03, +0.06$, respectively. This means that
the red region (between the contours $-0.06$ and $0.00$)
corresponds with the $1\sigma$ WMAP constraint \eqref{WMAPconstr},
while the yellow/green regions (near the axes) indicate values
that are too low, and the blue regions values that are too high.
The blue regions also correspond with a blue ($\tn>0$) spectrum.}
\end{figure}

An important result that can be seen from the equations in the
appendix, is that $\tgs_\mathrm{H}$ depends on the parameters $R$ and $x$
only. That means that the same is true for $\tn$. A contour plot of
$\tn$ as a function of $x$ and $(R-4)^{-1}$ is
given in figure~\ref{contn} (we use $(R-4)^{-1}$ for the ordinate
to avoid the curves getting squashed in the region
$R \downarrow 4$).
We conclude that there is a parameter region
(the red area) where a spectral index compatible with the
$1\sigma$ WMAP constraint \eqref{WMAPconstr} is produced. For $R<27/4$
(i.e.\ $(R-4)\inv > 4/11$) it is even possible to get $\tn>0$ (the
blue region). On the other hand, the amplitude does not just
depend on $R$ and $x$, but also on
$\ga_4$ explicitly, as can be seen from \eqref{amplitudeandslope}.

\begin{figure}
\begin{center}
\includegraphics[width=10cm]{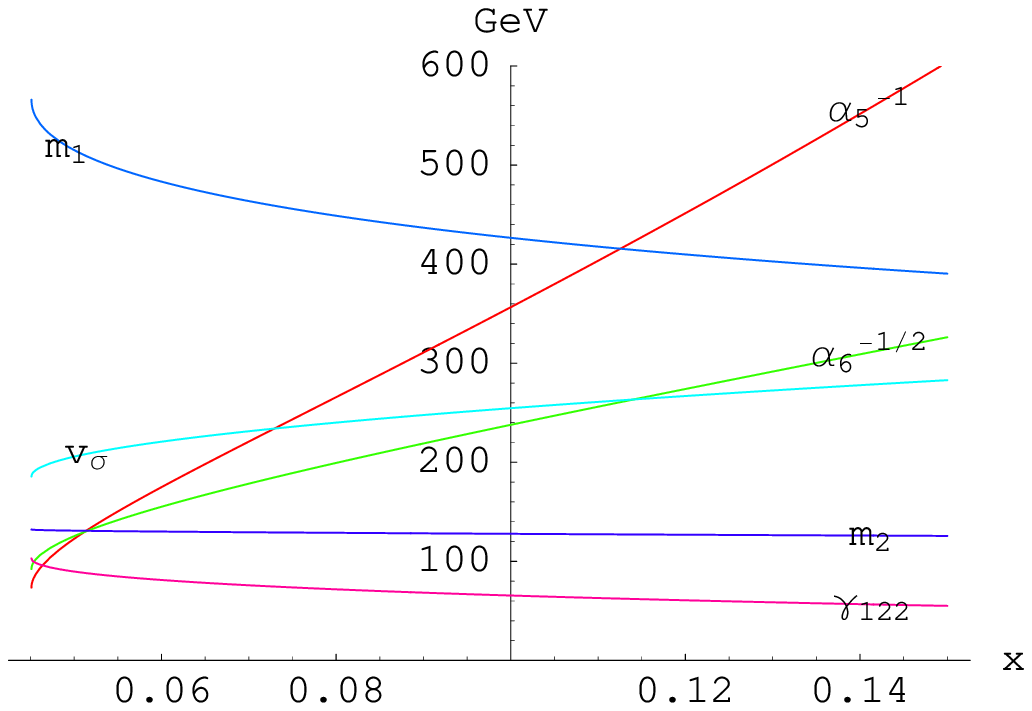}

\vspace{1cm}

\includegraphics[width=10cm]{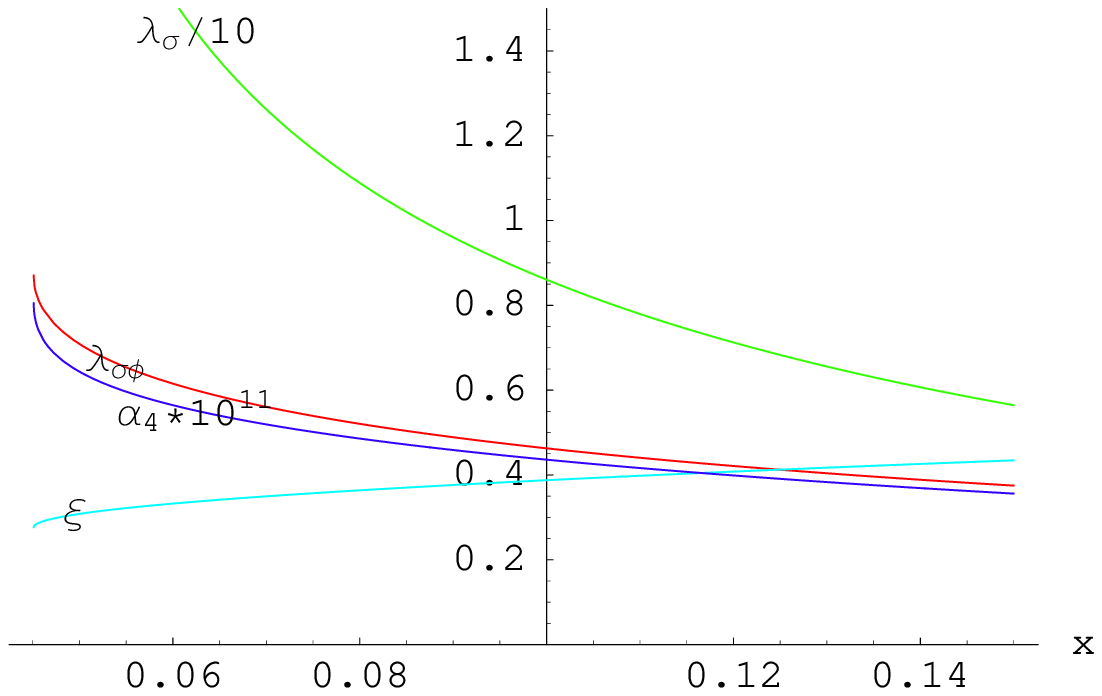}
\end{center}
\caption{\label{parsx}Variation of the dimensionful (top) and dimensionless
(bottom) parameters in the model described in
section~\ref{quartic} as a function of the one free parameter
$x$ when all CMB (with $\tn=-0.03$) and SM constraints are applied
(and choosing the smaller value for $R$).}
\end{figure}

Now we can apply all the constraints and determine the parameters
in the following way. First we determine
$x$ and $R$ from the spectral index constraint. Of course this
cannot be done uniquely; basically we are free to choose e.g.\ $x$
(within certain limits as indicated in figure~\ref{contn}) and
then $R$ is fixed by the constraint. Actually there are still two
possibilities for $R$, but it turns out that the smaller value
(corresponding to the upper branch in figure~\ref{contn}) leads to
smaller non-renormalizable couplings $\ga_5$ and $\ga_6$, and also
to a smaller dimensionless coupling $\gl_\gs$ (cf.\
(\ref{deflmsg})), which we consider more acceptable. Next
$\ga_4$ is fixed by the CMB amplitude constraint.
Finally, the other parameters are determined by the coupling to
the Higgs sector, in a way completely analogous to the treatment
in sections~\ref{Higgssector} and \ref{mixing}. Setting
$\tn=-0.03$, the central WMAP value, an explicit example with $x$
chosen to be $0.1$ is:
\be
\fl
x = 0.10:
\quad R=4.32,
\quad \ga_2^{1/2} = 1.30 \times 10^{-15}\, \mbox{GeV},
\quad \ga_4 = 4.36\times 10^{-12} ,
\non\\
\ga_5^{-1} = 357\, \mbox{GeV},
\quad \ga_6^{-1/2} = 238\, \mbox{GeV},
\quad \lmsp = 0.46,
\non\\
\gl_\gf = 0.33,
\quad \gl_\gs = 8.6,
\quad v_\gf = 246\, \mbox{GeV},
\quad v_\gs = 255\, \mbox{GeV},
\non\\
m_1 = 427\, \mbox{GeV},
\quad m_2 = 128\, \mbox{GeV},
\quad \xi=0.39,
\quad \gg_{122} = 65.5\, \mbox{GeV}.
\label{ex1}
\ee
For a more general picture see figure~\ref{parsx}, where all the parameters have
been plotted as a function of $x$ (and choosing the smaller value for $R$).
We see that as $x$ increases, the non-renormalizable mass scales
$\ga_5^{-1}$ and $\ga_6^{-1/2}$ also increase and that the (rather
large) inflaton self-coupling $\gl_\gs$ decreases. Since smaller
couplings are generically easier to deal with, the larger
$x$ values would be favoured, but we shall not pursue this aspect further.

\section{Loop corrections}
\label{loop}
Thus far we have assumed the potential $V$ to be an {\em
effective} potential that approximately describes some extension
of the SM, i.e.\ including all quantum effects. Nevertheless,
since the calculation of the spectral index is based on the growth
of quantum fluctuations during inflation, it is important to try
to ascertain that the back-reaction of these fluctuations on the
inflaton and on the SM is under control. After all, the $\gs$
values vary over some twelve orders of magnitude from the
inflationary to the electroweak domain. We therefore investigate
in this section one-loop corrections based on $V$.

Corrections on the effective potential in the inflationary domain
due to a Higgs loop have been investigated in
\cite{Copelandetal}, but not those coming from an inflaton
loop. There is reason for concern, because the four-point
self-coupling $\gl_\gs$ of the inflaton in the minimum of the
potential, defined in \eqref{deflmsg}, is rather large. For
example, $\gl_\gs = 8.6$ for the case (\ref{ex1}).
With such large couplings, computing loop `corrections' is
a hazardous endeavour and the calculations in this section are aimed
at obtaining insight rather that getting quantitative results.

At strong couplings the non-perturbative phenomenon of
`triviality' comes into play: as the cut-off used to define the
model is raised the renormalized couplings go down, and they even
vanish in the infinite cut-off limit. Larger couplings imply a
smaller maximum value of the cut-off, which is then interpreted as
a momentum scale where the model breaks down (also known as `the
Landau pole'). For a review of the application to the SM see
\cite{pdg,Sher}. Assuming that this phenomenon applies also here, we
shall tentatively use it to estimate the maximum cut-off from the
large inflaton self-couplings.

Another reason for expecting such a scale where the model breaks
down is the fact that it contains non-renormalizable couplings of
dimension larger than four. This situation may be compared with
effective pion models for QCD. These models are typically also
non-renormalizable (even non-polynomial) and in the first instance
only valid up to a few hundred MeV.\footnote{We recall also the
linear sigma model, in which the self-coupling is large,
$\gl = m_\gs^2/(2 f_\pi^2) \simeq 20$--$45$ for a sigma resonance
in the range 600--900 MeV.} Loop corrections
may then further extend their validity. A well-developed scheme is
chiral perturbation theory, in which all perturbative infinities
are removed by counter-terms, with new physical constants
parametrizing the corresponding finite parts
\cite{GasserLeutwyler,Weinberg}.

Before continuing in detail we should add the cautionary remark
that the usual equilibrium effective potential can in principle
not be applied blindly to systems out of equilibrium; we interpret
it as being only indicative of the back-reaction. More
sophisticated methods (e.g.\ Hartree or 2PI \cite{CormierHolman,Bergesreview})
are available, but they are a lot more complicated. In the
inflationary domain the one-loop potential is complex due to the
tree potential being unstable, and we shall concentrate on its
real part.

Consider the potential
\be
\fl
V^{(0)} = V_0
-\frac{1}{2}\ga_2\gs^2 +\frac{1}{4}\ga_4\gs^4
-\frac{1}{5}\ga_5\gs^5
+\frac{1}{6}\ga_6\gs^6
-\frac{1}{2}\lmsp\gs^2\gf^2 +
\frac{1}{2}\mu^2\gf^2 +
\frac{1}{4} \gl_\gf \gf^4,
\label{treepot}
\ee
which is to be identified with $V$ at tree level.
For simplicity we consider here only one real Higgs field to avoid
the complications of massless Goldstone bosons, which would be
absorbed by the W and Z bosons in the full SM. The one-loop
contribution to the effective potential is given by
(see e.g.\ \cite{Sher})
\be
V^{(1)} = \ge(m_1^2) + \ge(m_2^2) + {\rm c.t.}
\ee
where $\ge(m^2)$ is the ground-state energy density of a free
scalar field with mass $m$, c.t.\ denotes counter-terms and
$m_1^2$ and $m_2^2$ are the eigenvalues
\be
m_{1,2}^2 = \frac{1}{2}\left(m_{\gs}^2 + m_{\gf}^2 \pm
\sqrt{(m_{\gs}^2 -m_{\gf}^2)^2 + 4 m_{\gs\gf}^4}\right)
\label{m12sq}
\ee
of the field-dependent mass matrix
\be
m_{ab}^2 &=& V_{,\,ab}^{(0)},\qquad (a,b)\in \{\gs,\gf\},
\\
m_{\gs}^2 &\equiv& m_{\gs\gs}^2 =
-\ga_2+ 3\ga_4\gs^2-4\ga_5\gs^3+5\ga_6\gs^4- \lmsp \gf^2,
\\
m_{\gf}^2 &\equiv& m_{\gf\gf}^2 = \mu^2 -\lmsp \gs^2 + 3\gl_\gf \gf^2,
\qquad
m_{\gs\gf}^2 = -2\lmsp \gs\gf.
\ee
The ground-state energy density is given by
\be
\ge(m^2) =
\int_{|{\bf p}|<\gL}\frac{d^3 {\bf p}}{(2\pi)^3}\, \frac{1}{2}
\sqrt{m^2 + {\bf p}^2},
\label{c1}
\ee
where we used a spherical cut-off $\gL$ on the three-momenta.
Alternatively, we can use a four-dimensional Euclidean cut-off on
the usual log-determinant form
$\ln\det(m^2_{ab}+\partial^2)$, which leads to
\be
\ge(m^2) = \frac{1}{2}
\int_{|p|<\gL}\frac{d^4 p}{(2\pi)^4}\, \ln(m^2 + p^2).
\label{c2}
\ee
The first regularization (\ref{c1}) has the advantage that time is
kept real (and continuous), which is conceptually attractive for
systems out of equilibrium, whereas the regularization (\ref{c2})
can be applied only in equilibrium with imaginary and somewhat
fuzzy time. It turns out that both regularizations give very
similar results and for definiteness we shall continue with the
first one, (\ref{c1}).

Within the spirit of a polynomial parametrization, the
counter-terms are supposed to be polynomial in $\gs$ and
$\gf$. It is desirable that they are able to cancel all divergencies as
$\gL\rightarrow\infty$: the quartic ($\propto\gL^4$), quadratic
($\propto(m_1^2+m_2^2)\gL^2$) and logarithmic
($\propto((m_1^2)^2+(m_2^2)^2)\ln\gL$) ones.
This is possible, since the square root in (\ref{m12sq}) drops out of the sum
$m_1^2 + m_2^2$, and also
$(m_1^2)^2 + (m_2^2)^2 = (m_\gs^2)^2 + (m_\gf^2)^2 + 2 (m_{\gs\gf}^2)^2$
is a polynomial in $\gs$ and $\gf$. We define
\be
\ge_\mathrm{f}(m^2) &=& \ge(m^2)-\frac{1}{32\pi^2}\left(
2\gL^4+ 2m^2\gL^2-(m^2)^2 \ln\frac{2 e^{-1/4}\gL}{\nu}\right)
\label{c1f}
\\
&=&\frac{1}{64\pi^2}\, (m^2)^2 \ln\frac{m^2}{\nu^2} +
O(\gL^{-2}\ln\gL),
\label{c3}
\ee
where we introduced the renormalization scale $\nu$.
Dropping terms with negative powers of $\gL$,
the form (\ref{c2}) for $\ge$ reduces also to the form (\ref{c3})
(with
a different subtraction).
We shall continue with the full form (\ref{c1f}) with
$\ge(m^2)$ given in \eqref{c1}, without dropping negative powers of $\gL$.

The total potential at one loop is 
\be
\label{v0v1}
V = V^{(0)} + V^{(1)}.
\ee
At this stage we have used counter-terms corresponding to all the
terms in \eqref{treepot}
to cancel the divergent cut-off dependence, as well as a few more:
$\gs^7$, $\gs^8$,
$\gs^3 \gf^2$ and $\gs^4 \gf^2$.
The finite parts of these counter-terms have been assigned values
by the subtraction in (\ref{c1f}) (which is similar to the minimal
subtraction used in perturbative QCD calculations). However, we
still need at least some of these parameters to impose
renormalization conditions on the potential. A minimal set of
conditions is: ($i$) the real part of the expansion of $V$ in
$\gs$ around $\gs=0$, at $\gf=0$, coincides with the original
$V^{(0)}$ up to and including $\gs^5$; ($ii$) $V$ vanishes in its
absolute minimum at
$\gs=v_\gs$, $\gf=v_\gf$; and ($iii$) $v_\gf = 246$ GeV,
unchanged. Requirement ($i$) expresses our wish to keep the
potential unchanged in the inflationary domain, ($ii$) keeps the
cosmological constant zero and ($iii$) is needed for electroweak
phenomenology. Furthermore, in a sensible renormalized
perturbation scheme it is desirable that the particle masses in
loop diagrams are kept at the same values as in the tree graph
starting point, so we add to our minimal set of conditions: ($iv$)
$v_\gs$, ($v$) $m_{\gs}^2$ and ($vi$) $m_{\gf}^2$
are unchanged in the minimum of the potential. We can impose all
these conditions by using only counter-terms according to the
parameters listed in the tree potential (\ref{treepot}). It turns
out that
$m_{\gs\gf}^2$ ends up also very close to its tree value, so
the particle masses $m_1$ and $m_2$ are indeed practically
unchanged. The unusual combination of two classes of
renormalization conditions, ($i$) in the inflationary region and
($ii$)--($vi$) in the `today's physics' region, makes an
interesting problem.

Summarizing, the above renormalization conditions can be
implemented by counter-terms of the form
\be
\fl
\mbox{c.t.}=
-\frac{1}{32\pi^2}\left[ 4\gL^4+2(m_1^2+m_2^2)\gL^2 -
\left((m_1^2)^2 + (m_2^2)^2\right)\ln\frac{2 e^{-1/4}\gL}{\nu}\right]
\non\\
\fl \qquad\quad\mbox{}
- \sum_{k=0}^{5} \mbox{Re}
\left\{\frac{\partial^k}{\partial\gs^k}
\left[\ge_\mathrm{f}(m_\gs^2)+\ge_\mathrm{f}(m_\gf^2)\right]\right\}_{\gs=\gf=0}
\frac{\gs^k}{k!}
\label{ct}\\
\fl \qquad\quad\mbox{}
+\gd\ga_6\frac{\gs^6}{6}
-\gd\lmsp\frac{\gs^2\gf^2}{2}
+ \gd\mu^2 \frac{\gf^2}{2} + \gd\gl_\gf \frac{\gf^4}{4},
\non
\ee
which satisfies condition ($i$) (note that $m_1^2 =m_\gs^2$ and
$m_2^2=m_\gf^2$ in the inflationary region, since $\gf=0$).
The first thing to check is the coefficient of the $\gs^6$ term in
the expansion around $\gs=0$: it should not be large, as this
might spoil the neglect of such a term in the inflationary domain.
For the Higgs loop this has been done already in
\cite{Copelandetal}. Its contribution to the coefficient of
$\gs^6/6$ is
$-\lmsp^3/(32\pi^2\mu^2)$,
which for the case (\ref{ex1}) is only
$-3.1\times 10^{-8}$~GeV$^{-2}$,
even (much) smaller (in absolute value) than
$\ga_6= 1.8\times 10^{-5}$~GeV$^{-2}$.
For the inflaton loop the effects are even smaller because the
effective mass $m_\gs$ is tiny in the inflationary region;
although $\ln(m_\gs^2/\nu^2)$ is the logarithm of a very small
number, the $(m_{\gs}^2)^2$ in front of it makes it negligible.
Since the terms with different powers of $\gs$ up to and including
$\gs^5$ are by construction all approximately equal during
inflation, a rough estimate of the relative importance of the
inflaton loop corrections is given by
$\ga_2^2 \ln (\ga_2/\nu^2) / (64\pi^2 \ga_2 \gs_\mathrm{H}^2) = \cO(10^{-12})$.

We now impose the renormalization conditions ($ii$)--($vi$) in
the minimum of the potential. These are five linear equations for
the four parameters of the counter-terms in the third line of
(\ref{ct}), plus the renormalization-scale parameter $\ln(\nu)$.
Using as an example the parameter set (\ref{ex1}) and choosing
$\gL=\infty$ for a start, results in
\be
\fl
\gd\ga_6/\ga_6=0.10,
\quad
\gd\gl_{\gs\gf}/\gl_{\gs\gf}=-0.011,
\quad
\gd\mu^2/\mu^2 = -0.058,
\quad
\gd\gl_\gf/\gl_\gf = 0.026,
\non\\
\fl
\nu=939\, \mbox{GeV},
\quad
m_{\gs\gf}^{2(0+1)}/m_{\gs\gf}^{2(0)}=0.95,
\quad
\gl_{\gs}^{(0+1)}/\gl_{\gs}^{(0)}=2.07,
\label{out1}
\ee
where (0) indicates the tree graph value and (0+1) the value derived from the 
full potential (\ref{v0v1}). The renormalization
scale $\nu$ seems to be somewhat large, but small changes in
$\ln(\nu)$ are of course amplified in $\nu$ itself. The finite
counter-terms look reasonably small. The mixing mass
$m_{\gs\gf}$ is also close to the tree graph value,
in accordance with our desire to keep the particle masses in the
loop contributions close to the tree values.
The inflaton self-coupling
$\gl_{\gs}$ has increased by an unpleasant factor of about two,
which seems to make the calculation untrustworthy. Another way of
treating the renormalization to deal with this is discussed in
appendix~B.

Since we have removed the divergent cut-off dependence with the
counter-terms, the renormalized effective potential cannot give us
information on the scale where the model is expected to break
down. For this we look at the bare coupling defined in terms of
the bare potential $V^{(0)} +\mbox{c.t.}$, by
\be
\gl_\gs^{(\rm b)} =(1/6) (V^{(0)} +\mbox{c.t.})_{,\,\gs\gs\gs\gs},
\ee
evaluated at its absolute minimum. In a simple $\gf^4$ model with
one real scalar field the relation between the bare and
renormalized coupling is qualitatively well described by the
one-loop renormalization-group relation (see e.g.\
\cite{Sher,Heller})
\be
\gl^{(\rm b)} = \frac{\gl}{1-\gb \gl t}= \gl + \gl^2 \gb t +
\cdots,
\qquad
t= \ln(\gL/m) + {\rm constant},
\label{Lp}
\ee
with $\gb= 9/(8\pi^2)$ the one-loop beta-function coefficient. The
bare coupling $\gl^{(\rm b)}$ diverges when the denominator
vanishes at the position of the `Landau pole', and numerical
simulations have indicated that this gives an order of magnitude
estimate of the limiting $\gL$ (modulo the uncertainties related
to methods of regularization and factors like $2e^{-1/4}$ in
\eqref{c1f}; for a review see e.g.\ \cite{Heller}). As
can be seen from the expansion in (\ref{Lp}), the one-loop value
of the bare coupling equals twice the tree value at the pole, and
we shall use this for an estimate of the maximal cut-off $\gL_{\rm
max}$:
\be
\gl_\gs^{(\rm b)}/\gl_\gs = 2 \quad \Rightarrow \quad \gL = \gL_{\rm max}.
\label{maxcutoff}
\ee
Using the results from the renormalization procedure in
(\ref{out1}) we find that the maximum value is already reached
near
$\gL\simeq 600$~GeV, for which $\gl_\gs = 12.8$ (and $\nu = 702$ GeV).

The loop calculation in this section has given evidence that the
flatness of the inflationary part of the potential can be kept
consistent with quantum corrections. It appears that the model has
to already break down at a fairly low scale. For the example
(\ref{ex1}) this could be around $700$~GeV or perhaps as low as
$300$~GeV (see appendix~B), but for larger $x$ values this scale
slowly increases. Above this scale new physical input is needed.

\section{Conclusion}
\label{concl}

In this paper we investigated the
implications of the WMAP results for low-scale inflation and found that there
are severe constraints.  This is essentially because the proximity of the
observed scalar spectral index $\tn\equiv n-1$ to zero is hard to reconcile
with the  small number of e-folds between horizon crossing of the observable
scales and  the end of inflation in these models.  Working in the context of a
phenomenological electroweak-scale inflation model that allows for tachyonic
electroweak baryogenesis and consists of one additional scalar field $\gs$
coupled to the Standard Model Higgs, we were led to further constraints on the
inflaton--Higgs potential. However, we found that there is a range of
parameters compatible with a spectral index close to (and even larger than)
zero in this model.

The polynomial approximation together with all the constraints led
to the conclusion that we need
a $\gs^5$ term in the potential during inflation (as in \cite{Copelandetal}). 
In addition
$\gs^2$ and $\gs^4$ terms are needed, and (or instead of the
$\gs^4$) there might be a $\gs^3$ term, but no powers higher than
5 can be present during inflation. The appearance of an odd power
($\gs^5$) implies a local minimum at small negative values of the
inflaton field ($\gs=0$ being the value where the potential has a
local maximum). A universe with initial conditions in this
negative region would classically never stop inflating, which is
why we assumed a small positive initial condition for the inflaton
field. However, quantum tunnelling might very well make the case of
a negative initial condition viable as well. The use of a
polynomial approximation is quite natural for the inflationary
region of the potential where the inflaton field is small, but it
seems somewhat artificial in the large field region where the
Higgs field comes into play. The odd and non-renormalizable power
$\gs^5$ in the potential appears to be the price we have to pay
for keeping the number of parameters limited. It is not excluded
of course that the model with its non-symmetric potential and
non-renormalizable couplings can be embedded satisfactorily in a
model with more symmetry that is also renormalizable, e.g.\ a
supersymmetric extension of the Standard Model.

In the one-loop calculation we found that the very different
renormalization conditions in the inflationary and electroweak
regimes did not lead to unresolvable conflicts. In our results
quantum corrections do not disrupt the required flatness of the
potential in the inflationary region. The non-renormalizable
couplings and also the relatively strong inflaton self-coupling
suggest a breakdown of the model already occurring at a fairly low scale,
perhaps below 1 TeV, depending on the choice of parameters.

We conclude with the important remark that the phenomenological
model arrived at here can be falsified experimentally through its
conspicuous generic feature: the existence of two (and only two)
particle species with zero spin and masses around the electroweak
scale, and couplings to the rest of the SM equal to that of the
Higgs up to factors related to the mixing angle.

\ack
This work is supported by FOM/NWO and by PPARC Astronomy Rolling
Grant PPA/G/O/2001/00476.

\appendix
\section{Some analytical results for the full model with a quartic term}

Starting from equations \eqref{Vinfl} and \eqref{defsRx} in
section~\ref{quartic}, we find the following slow-roll field
equation:
\be
\d N = \frac{R}{x} \, \frac{\d\tgs}{R \tgs - 3 \tgs^3 + \tgs^4}.
\ee
Integrating it from horizon crossing (subscript H) to the end
of inflation (subscript e), we obtain for $R>4$ (with $N_k
\equiv N_\mathrm{e}-N_\mathrm{H}$):
\be
\fl
N_k = \frac{1}{3x} \Biggl [
\frac{2+Q+2Q^2}{1+Q+Q^2} \ln \lh \frac{\tgs-1+Q+Q^{-1}}
{\sqrt{\tgs^2-(2+Q+Q^{-1})\tgs + Q+Q^{-1}+Q^2+Q^{-2}}} \rh
\non\\
+ 3 \ln \lh \frac{\tgs}{\tgs-1+Q+Q^{-1}} \rh
+ \sqrt{3} \, \frac{Q(Q+1)}{Q^3-1} \, F(\tgs)
\Biggr ]^{\tgs=\tgs_\mathrm{e}}_{\tgs=\tgs_\mathrm{H}},
\ee
with
\be
F(\tgs) \equiv \left \{ \begin{array}{ll}
\arctan \lh \frac{\frac{1}{2}\sqrt{3} (Q-Q^{-1})}
{1+\frac{1}{2}Q+\frac{1}{2}Q^{-1}-\tgs} \rh
& \mbox{for $\tgs < 1+\frac{1}{2}Q+\frac{1}{2}Q^{-1}$}\\
\arctan \lh \frac{\frac{1}{2}\sqrt{3} (Q-Q^{-1})}
{1+\frac{1}{2}Q+\frac{1}{2}Q^{-1}-\tgs} \rh + \pi
& \mbox{for $\tgs > 1+\frac{1}{2}Q+\frac{1}{2}Q^{-1}$}
\end{array} \right.
\ee
and
\be
Q^3 \equiv \frac{1}{2} \lh R-2+\sqrt{R} \, \sqrt{R-4} \rh
\qquad\Leftrightarrow\qquad
R = 2+Q^3+Q^{-3}.
\ee
The field value at the end of inflation that follows from the condition
$\tget=1$ is for $x<1$ given by
\be
\fl
\tgs_\mathrm{e} = \frac{3}{4} \lh 1 + S + S^{-1} \rh,
\quad
S^3 \equiv \frac{8}{27} \lh R \, \frac{1-x}{x} + \frac{27}{8}
- \sqrt{R \, \frac{1-x}{x}} \, \sqrt{R \, \frac{1-x}{x} + \frac{27}{4}} \rh.
\ee

\section{Another renormalization treatment}

We found in \eqref{out1} that the already large inflaton
self-coupling
$\gl_\gs$ increased by another factor of two when applying the renormalization
conditions. Of course, {\it de facto}, $\gl_\gs$ is a derived
coupling: it depends on the other parameters of the model and we
may just have to accept how it turns out. In principle, further
experimental information is needed to be able to pin down the
non-renormalizable couplings beyond $\ga_5$ and $\ga_6$, similar
to what is done in chiral perturbation theory. To continue, we can
pretend that our tree level values for couplings such as
$\gl_\gs$ are actually phenomenologically correct, i.e.\ consider
them as experimental input. Alternatively (but leading to the same
treatment) we can work from the point of view presented at the end
of section~\ref{mixing} that
$\gl_\gs$ is a primary coupling and that the other parameters have
to be chosen accordingly. (At this point we have assigned values
to the non-renormalizable couplings of the $\gs^7$, $\gs^8$,
$\gs^3 \gf^2$ and $\gs^4 \gf^2$ terms in a somewhat arbitrary way by the
subtraction in (\ref{c1f}) and the value of the renormalization scale $\nu$.)
So we continue this exploration by imposing that in addition to $v_\gs$,
$v_\gf$ and the particle masses
$m^2_\gs$, $m^2_\gf$ and $m^2_{\gs\gf}$,
also the tree graph value of $\gl_\gs$ is to remain unchanged, and
that the other parameters have to be chosen accordingly.

We add to our renormalization conditions: ($vii$)
$m^2_{\gs\gf}=m^{2(0)}_{\gs\gf}$ and ($viii$) $\gl_\gs =
\gl_\gs^{(0)}$. Since $\gl_\gs$ is naturally controlled by the
$\ga$ parameters, we add to (\ref{ct}) the finite counter-terms
\be
+ \, \gd\ga_7\frac{\gs^7}{7}+\gd\ga_8\frac{\gs^8}{8},
\ee
still keeping the coefficients of $\gs^3\gf^2$ and $\gs^4 \gf^2$
set by the value of $\nu$. The result of this exercise is given by
\be
\fl
\gd\ga_6/\ga_6=-0.75,
\quad
\gd\gl_{\gs\gf}/\gl_{\gs\gf}=0.15,
\quad
\gd\mu^2/\mu^2 = 0.34,
\quad
\gd\gl_\gf/\gl_\gf = 0.092,
\non\\
\fl
\nu = 3987\, \mbox{GeV},
\quad
\gd\ga_7^{-1/3}=217\, \mbox{GeV},
\quad
(-\gd\ga_8)^{-1/4}= 363\, \mbox{GeV}.
\label{out2}
\ee
We see that the scale of $\gd\ga_7$ and $\gd\ga_8$ is similar to
that of $\ga_5$ and $\ga_6$, and the other counter-term parameters
are larger than before in (\ref{out1}) but still reasonable,
except that $\nu$ has slipped to a rather large value. We could
fix $\nu$ to, say, 400 GeV and use one of the $\gs^k \gf^2$
couplings in its place, but then the resulting counter-term
parameters for the other couplings are larger. Moreover, we prefer
to keep $\nu$ variable since it adjusts itself naturally in such a
way that the loop correction vanishes as
$\gL\rightarrow 0$.

When we try to determine the maximal cut-off in this case using the
estimate \eqref{maxcutoff}, we find that this criterion cannot be
applied meaningfully because the bare potential quickly becomes
unstable (no lower bound) as
$\gL$ is increased from zero. This happens already for $\gL$
between 200 and 300 GeV (where $\nu$ has come down to $\approx
300$ GeV). Such a low value is also suggested by the fact that
$250$--$350$ GeV is
the scale of the non-renormalizable couplings
$\ga_5$ and $\ga_6$ in the example (\ref{ex1}).

\end{document}